\documentclass[apjl]{emulateapj}  
\usepackage{graphics}
\usepackage{amsmath}
\usepackage{bm}
\usepackage{rotating}
\usepackage[normalem]{ulem}

\shorttitle{CO in the Cold Debris of SN~1987A}
\shortauthors{Kamenetzky et al.}
\newcommand{\ms}{M$_{\odot}$}

\begin{document}

\title{Carbon Monoxide in the Cold Debris of Supernova 1987A}

\author{J. Kamenetzky\altaffilmark{1}, R. McCray\altaffilmark{1}, R. Indebetouw\altaffilmark{2,3}, 
M. J. Barlow\altaffilmark{4}, M. Matsuura\altaffilmark{4},
M. Baes\altaffilmark{5}, J. A. D. L. Blommaert\altaffilmark{6, 7}, A. Bolatto\altaffilmark{8}, 
L. Decin\altaffilmark{6}, L. Dunne\altaffilmark{9}, C. Fransson\altaffilmark{10},
J. Glenn\altaffilmark{1}, H. L. Gomez\altaffilmark{11}, M.A.T. Groenewegen\altaffilmark{12}, 
R. Hopwood\altaffilmark{13}
R.P. Kirshner\altaffilmark{14},
M. Lakicevic\altaffilmark{15},
J. Marcaide\altaffilmark{16}, I. Marti-Vidal\altaffilmark{17}, M. Meixner\altaffilmark{18}, 
P. Royer\altaffilmark{6}, 
A. Soderberg\altaffilmark{14}, G. Sonneborn\altaffilmark{19}, L. Staveley-Smith\altaffilmark{20,21},
B. M. Swinyard\altaffilmark{4}, G. Van de Steene\altaffilmark{12}, P.A.M. van Hoof\altaffilmark{12}, J. Th. van Loon\altaffilmark{15}, J. Yates\altaffilmark{4}, 
G. Zanardo\altaffilmark{20}
}
 
\altaffiltext{1}{Department of Astrophysical and Planetary Sciences, University of Colorado at Boulder, UCB 391, Boulder, CO 80309, USA} 
\altaffiltext{2}{Department of Astronomy, University of Virginia, PO Box 3818, Charlottesville, VA, 22903, USA} 
\altaffiltext{3}{National Radio Astronomy Observatory, 520 Edgemont Road, Charlottesville, VA 22903, USA} 
\altaffiltext{4}{Department of Physics and Astronomy, University College London, Gower St., London WC1E 6BT, UK} 
\altaffiltext{5}{Sterrenkundig Observatorium, Universiteit Gent, Krijgslaan 281 S9, B-9000 Gent, Belgium} 
\altaffiltext{6}{Instituut voor Sterrenkunde, KU Leuven, Celestijnenlaan 200D BUS 2401, 2001 Leuven, Belgium}
\altaffiltext{7}{Department of Physics and Astrophysics, Vrije Universiteit Brussel, Pleinlaan 2, 1050 Brussels, Belgium}
\altaffiltext{8}{Department of Astronomy, University of Maryland, College Park, MD 20742, USA}
\altaffiltext{9}{Department of Physics and Astronomy, University of Canterbury, Private Bag 4800, Christchurch, 8410, NZ}
\altaffiltext{10}{Department of Astronomy, The Oskar Klein Centre, Stockholm University, AlbaNova, SE--106~91 Stockholm, Sweden}
\altaffiltext{11}{School of Physics and Astronomy, Cardiff University, The Parade, Cardiff CF24 3AA, UK} 
\altaffiltext{12}{Royal Observatory of Belgium, Ringlaan 3, 1180 Brussels, Belgium}  
\altaffiltext{13}{Physics Department, Imperial College London, London SW7 2AZ, UK} 
\altaffiltext{14}{Harvard College Observatory, 60 Garden Street, Cambridge, MA 02138, USA}
\altaffiltext{15}{Lennard-Jones Laboratories, Keele University, ST5 5BG, UK} 
\altaffiltext{16}{Universidad de Valencia, C/ Dr. Moliner 50, 46100, Burjassot, Spain}
\altaffiltext{17}{Onsala Space Observatory, SE-439 92 Onsala, Sweden} 
\altaffiltext{18}{Space Telescope Science Institute, 3700 San Martin Drive, Baltimore, MD 21218, USA}
\altaffiltext{19}{NASA Goddard Space Flight Center, 8800 Greenbelt Rd, Greenbelt, MD 20771, USA}
\altaffiltext{20}{International Centre for Radio Astronomy Research (ICRAR), University of Western Australia, Crawley, WA 6009, Australia} 
\altaffiltext{21}{ARC CEntre of Excellence for All-sky Astrophysics (CAASTRO), University of Western Australia, Crawley, WA 6009, Australia}

\begin{abstract}
We report spectroscopic and imaging observations of rotational transitions of cold CO and SiO in the ejecta of SN1987A, the first such emission detected in a supernova remnant.  In addition to line luminosities for the CO J=1-0, 2-1, 6-5, and 7-6 transitions, we present upper limits for all other transitions up to J=13-12, collectively measured from the Atacama Large Millimeter Array (ALMA), the Atacama Pathfinder EXperiment (APEX), and the {\it Herschel} Spectral and Photometric Imaging REceiver (SPIRE).  Simple models show the lines are emitted from at least 0.01 \ms\ of CO at a temperature $> 14$ K, confined within at most 35\% of a spherical volume expanding at $\sim$ 2000 km s$^{-1}$.  Moreover, we locate the emission within 1\arcsec\ of the central debris.  These observations, along with a partial observation of SiO, confirm the presence of cold molecular gas within supernova remnants and provide insight into the physical conditions and chemical processes in the ejecta.  Furthermore, we demonstrate the powerful new window into supernova ejecta offered by submillimeter observations.
\end{abstract}

\keywords{ISM: supernova remnants - supernovae: individual (SN1987A)}

\section{Introduction}

The ejecta of supernovae (SN) encode important information about the SN explosion dynamics, 
the nucleosynthetic yield, and the formation of interstellar dust and molecules.  
At optical and X-ray wavelengths, the dominant feature of SN~1987A is a 
rapidly brightening 1.66 $\pm$ 0.03\arcsec\ \citep{Panagia1991} ring of ``hot spots'' where the blast wave is 
colliding with the innermost of three circumstellar rings \citep{McCray2007}.  
In the radio, this ring can be seen as two bright lobes of emission \citep{Ng2008, Zanardo2013}. 
Inside this ring, optical and near-infrared spectra showed that the debris from the progenitor's 
stellar interior is clumped into chemically distinct fragments, originating from the 
nucleosynthesis that took place before and during the supernova explosion.  
As the remnant cooled, dust formed and began to obscure the inner ejecta from view at optical wavelengths within the first few years.  
It was recently estimated that the ejecta of SN1987A contains 0.4-0.7 \ms\ of dust at a temperature of ~20 K  \citep{Matsuura2011}.
Fortunately, observations at far-infrared (FIR) to millimeter wavelengths pierce through the dust, providing a window 
into the cool interior \citep{Biermann1992,Lakicevic2011}.

A small mass ($10^{-3}$ \ms) of hot, vibrationally excited CO was seen 192 days after the 
explosion  \citep{Liu1992}, but the emission faded below detection limits after 
$\sim$600 days, when the gas became too cool to excite vibrational transitions \citep[700 K at 800 days]{Liu1995}.  
Theoretical models predict that CO formation would continue 
after that time, yielding $\sim$ 0.1 \ms\ of CO \citep{Cherchneff2009,Cherchneff2011} by 1000 days.  
Such a mass of CO would have important implications for the evolution of the ejecta, 
as discussed in \citet{Cherchneff2011}: first, dust synthesis can be limited by the depletion 
of elements by molecules, and second, CO is a powerful coolant, and a cooler
environment will favor more efficient dust and molecule formation.  

The ejecta of at least 8 other supernove have been observed in the vibrational CO fundamental (4.65 $\mu$m)
or first overtone (2.3 $\mu$m) bands, all within
the first few years after their explosion \citep[see references listed within their Table 1]{Cherchneff2011}.  
A notable addition is the 300 year old Cassiopeia~A remnant, where \citet{Rho2012} detected vibrationally excited $^{12}$CO 
with a mass of $\sim10^{-6}$ \ms\ and temperature as high as 900-1400~K.    
Until now no direct evidence (via {\it rotational} molecular emission) had 
been found for cold ($< 900$ K) molecular gas within supernova remnants, and no measurement 
made between $\sim 2$ and 300 years post-explosion.  With the Atacama Large Millimeter Array (ALMA),
we have now imaged the ejecta of SN1987A and, for the first time, detected rotational transitions of CO inside a supernova remnant.

\section{Observations}\label{sec:obs}

\subsection{ALMA}

SN~1987A was observed with ALMA at wavelengths of 2.6 mm (Band 3) and 1.3 
mm (Band 6) between April and August 2012.  During the 2012 Early Science 
period, ALMA had only a fraction of its final spatial resolution and 
sensitivity, but could already 
obtain synthesized beam spatial resolutions of $\sim$ 1.6 $\times$ 1.5\arcsec\ and 
0.6 $\times$ 0.5\arcsec, respectively.  Observation dates and parameters are listed in Table~\ref{obsparm}.
Both bands were observed
using Frequency Division Mode (FDM), i.e. 488 kHz channels, to enable detection of any
narrow emission lines (none were detected), and later averaged to 50
km s$^{-1}$ spectral resolution (corresponding to one hundred sixty 19 MHz channels in Band 3).  The bandwidth per spectral window was 1.875 GHz.
All observations used quasars J0538-440 and J0637-752 as bandpass and phase calibrators,respectively.    
Callisto was observed as an absolute flux calibrator; 
the absolute flux scale was then applied to the X352 and X1c blocks by equating the flux densities of the two quasars to the flux densities calibrated against a solar system object.
System temperature and precipitable water vapor (PWV) calibrations were applied before standard bandpass and complex gain
(phase) calibration.  
In the synthesis imaging, the robust parameter was set to 0, 
midway between natural and uniform weighting \citep{Briggs1995}.

\begin{deluxetable*}{lccccccccc}[t]
\tablecaption{ALMA Observational Parameters\label{obsparm}}
\tabletypesize{\scriptsize}
\tablewidth{\textwidth}
\tablehead{
\colhead{UID (uid://)} & \colhead{Band} & \colhead{Date} & \colhead{{\# Ant}\tablenotemark{a}} & \colhead{PWV} & \colhead{Int\tablenotemark{b}} &  \colhead{RMS\tablenotemark{c}}  & \colhead{BL$_{\rm RMS}$\tablenotemark{d}} & \colhead{BL$_{\rm max}$\tablenotemark{d}} \\
 &  & \colhead{(2012)} &  & \colhead{[mm]} & \colhead{[min]} & \colhead{[mJy bm$^{-1}$]}  & \colhead{[m]} & \colhead{[m]}
}
\startdata
A002/X3c5ee0/X24b       & 3\tablenotemark{e}  & 05 Apr & 16/18    & 3.1 & 25   &  0.13 & 220        & 403 \\
A002/X3c7a84/X1c        & 3  & 06 Apr & 14/17    & 3.7 & 25   &  0.12 & 225        & 402 \\
A002/X3c8f66/X352       & 6\tablenotemark{f}  & 07 Apr & 14/17    & 3.7 & 25   &  0.20 & 225        & 402 \\
A002/X45f1dd/Xd13       & 6  & 15 Jul & 16/19    & 1.3 & 12   &  0.14 & 213        & 402 \\
A002/X494155/X8be       & 6  & 10 Aug & 22/23    & 1.4 & 25   &  0.09 & 194        & 402 
\enddata
\tablenotetext{a}{Effective number of antennae after flagging / the total number in the array at the time.}
\tablenotetext{b}{Time on the science target, not including calibration overhead.}
\tablenotetext{c}{RMS (root mean square) noise is measured over a width of 100 km s$^{-1}$}
\tablenotetext{d}{Baseline.}
\tablenotetext{e}{100.1--103.9 and 112.2-118.8 GHz.}
\tablenotetext{f}{212.5--216.3 and 228.7--232.4 GHz.}
\end{deluxetable*}

\begin{figure}
\includegraphics[width=\columnwidth]{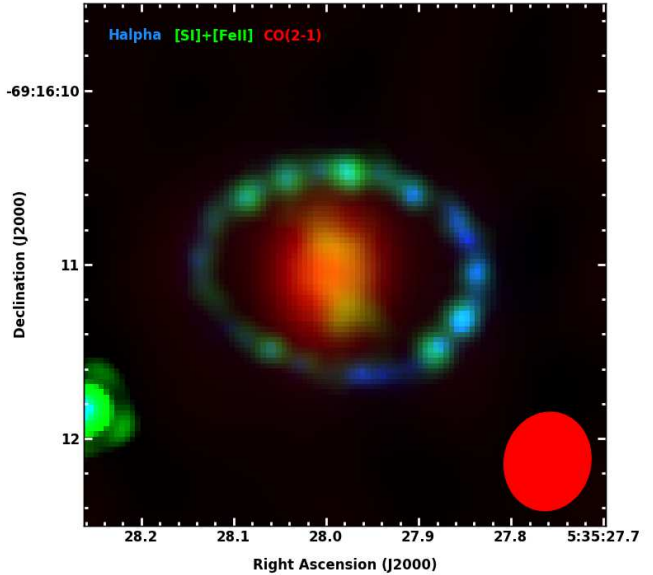} 
\caption{A color composite image of SN~1987A. The unresolved $^{12}$CO 2-1 line emission 
detected by ALMA is shown in red, and the red ellipse in the corner is the synthesized beam.  Also shown are the H$\alpha$ emission 
(blue) and [Si~{\sc i}]+[Fe~{\sc ii}] 1.644 $\mu$m emission (green in the 
ring; yellow in the ejecta) observed with the Hubble Space Telescope \citep{Larsson2013}.
\label{fig:almamap}}
\end{figure}

\begin{figure}
\includegraphics[width=\columnwidth]{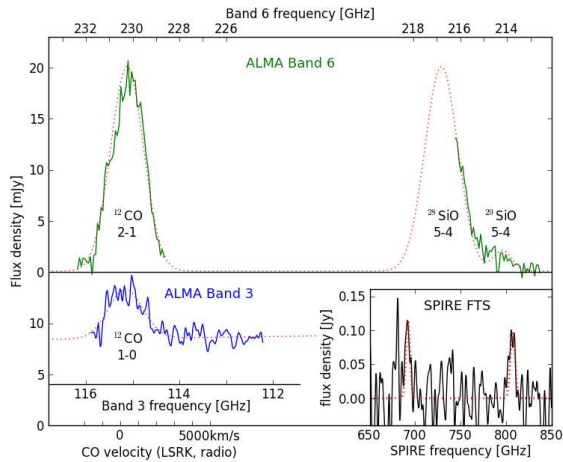}
\caption{ALMA spectra at the center of SN1987A.  The 2.6mm Band 3 (bottom left) and 
the 1.3mm Band 6 (top) are plotted with a common velocity axis calculated for 
the $^{12}$CO 1-0 and 2-1 lines, respectively.  The dotted 
lines are the best-fit Gaussians to the $^{12}$CO lines, $^{28}$SiO and 
$^{29}$SiO and a continuum with synchrotron spectral index of $-0.8$.  
A zoomed-in portion of the continuum-subtracted SPIRE spectrum is shown in the bottom right, with 
the best-fit Gaussians of the J=6-5 and J=7-6 lines as dotted lines.
\label{fig:almaspec}}
\end{figure}

We have detected emission from the rotational lines of $^{12}$CO J=1-0 (115.3 GHz, 2.6mm) 
and J=2-1 (230.6 GHz, 1.3mm), as well as the red wing of SiO J=5-4 at 
217.1 GHz   Figure~\ref{fig:almamap} is a color composite image showing 
the $^{12}$CO 2-1 emission detected by ALMA (red) and other emission (H$\alpha$ and [Si~{\sc i}]+[Fe~{\sc ii}]) for comparison.
The origin of the $^{12}$CO J=2-1 emission, marginally larger in extent than the beam (unresolved), 
is shown here to be less than 1\arcsec\ in extent and located at the center of the debris.
The spectrum shown in Figure~\ref{fig:almaspec} of the center of the remnant includes 
some continuum emission.  However, consistent with observations by \citet{Lakicevic2012b}, 
this can be identified with the synchrotron emission from the ring. 
The line profiles are broad (FWHM $\sim 2200$ km s$^{-1}$, Table \ref{table:fit}), 
consistent with the near-infrared [Si~{\sc i}]+[Fe~{\sc i}] lines \citep{Kjaer2010} and $^{12}$CO rotational-vibrational models \citep{Liu1992},
confirming an origin from the expanding ejecta (and a constant expansion velocity over time).
The lines are narrower than 
those (4,500 to 11,000 km s$^{-1}$) of ejecta metal lines observed in the visible \citep{Fransson2013}, 
indicating that the $^{12}$CO emission originates from a more centrally-condensed source.  

Although our intention with Band 6 was to observe the $^{12}$CO J=2-1 line, the red wing of the $^{28}$SiO J=5-4 line at 
217.11 GHz appeared in the upper sideband.  If it has the same profile as $^{12}$CO J=2-1, then the amplitude would be 20 $\pm$ 5 mJy.  
Because we do not measure the full flux in the line, we do not conduct further analysis 
at this time.  We see a faint hint of the isotopic $^{29}$SiO line at 214.39 GHz (at about 1/10 the estimated amplitude of $^{28}$ SiO, though narrower), but with marginal statistical significance.

\begin{deluxetable*}{ l c c c c c }
\tablecaption{Fitted Line Parameters.\label{table:fit}}
\tablewidth{\textwidth}
\tablehead{
\colhead{Line}  & \colhead{FWHM} & \colhead{Central Velocity} & \colhead{Peak Intensity} & \colhead{Integrated Flux}  & \colhead{Luminosity\tablenotemark{a}}\\ 
    & \colhead{[km s$^{-1}$]}  & \colhead{[km s$^{-1}$]} & \colhead{[mJy/beam]} & \colhead{[Jy km s$^{-1}$]} & \colhead{[10$^{31}$ erg s$^{-1}$]}
}
\startdata
        $^{12}$CO 1-0 & 2270 $\pm$ 190 & 300 $\pm$ 100    & 4.8 $\pm$ 0.3  & 11.6 $\pm$ 1.2   & 1.33 $\pm$ 0.14          \\ 
        $^{12}$CO 2-1 & 2150 $\pm$ 50  & 390 $\pm$ 20     & 20.0 $\pm$ 0.7 & 46.0 $\pm$ 1.9   & 10.5 $\pm$ 0.4        
\enddata
\tablecomments{Uncertainties are one standard deviation. 
In subsequent analysis we added in quadrature a 10\% continuum subtraction 
error and a 15\% (10\%) absolute flux calibration error for $^{12}$CO 1-0 (2-1), 
leading to total line luminosity uncertainties $0.28 \times 10^{31}$ and 
$1.5 \times 10^{31}$ erg s$^{-1}$, respectively.}
\tablenotetext{a}{Luminosities are calculated using a distance of 50 kpc throughout this work \citep{Panagia1999}.}
\end{deluxetable*}

\subsection{APEX}

We also observed the $^{12}$CO J=3-2 line of SN 1987A using the Atacama Pathfinder Experiment (APEX) in Chile (European Southern Observatory proposal ID: 090.D-0515(A)).
We used the SHeFI receiver \citep{Nystrom2009} at the central frequency of 345.796 GHz, and upper side band and wobbler beam switching mode with two different chopping distances of 40 and 45\arcsec.  
The back end system has 8192 $\times$ 2 channels of 61 kHz width.  The two bands are 4 GHz wide each (1 GHz overlapping).  The system temperature was typically 200 K.
We chose 1.5 hours (of the most stable sky conditions) of the total 4.5 hours of data acquired over four days (2012 October 6, 8, 10 and 16, PWV from 0.2--0.7 mm).
The focus was checked with Jupiter, {\it o} Cet and R Dor; the pointing was regularly checked and updated on the nearby pointing source R Dor.   
The antenna temperature was converted to main beam temperature and then to flux in Jy using 41 Jy K$^{-1}$ (for the 17\arcsec\ beam). We subtracted the 
baseline after combining all the spectra using the CLASS software and obtained an upper limit on the $^{12}$CO J=3-2 
line of 94 mJy (5.2 $\times 10^{32}$ erg s$^{-1}$).

\subsection{SPIRE}

The $^{12}$CO J=4-3 to J=13-12 transitions are encompassed by the SPIRE \citep{Griffin2010}  Fourier Transform 
Spectrometer (FTS, 447 to 1550 GHz) onboard the {\it Herschel} Space Observatory \citep{Pilbratt2010}.  
SN1987A was observed by SPIRE on 2012 June 12 for $\sim$ 4 hours (Observation ID
1342246989, proposal GT2\_mbarlow\_1) as part of the MESS program \citep{Groenewegen2011} and reduced in HIPE v11.
We detected two lines, J=7-6 and J=6-5, at $(7.0 \pm 2.5$) and ($8.0 \pm 2.0) \times 10^{-18}$ W m$^{-2}$.
In luminosity, they are (2.01 $\pm 0.8$) and ($2.4 \pm 0.6$) $\times 10^{33}$ erg s$^{-1}$, respectively.
These lines were fit with a FWHM of 2300 km s$^{-1}$\ (Figure \ref{fig:almaspec}), consistent in origin with the emission measured by ALMA.  Those two CO lines fall in the lowest noise region of the entire band (the sensitivity is wavelength dependent), though we were also able to derive upper limits of 1.3-5.2 $\times 10^{-17}$ W m$^{-2}$ for the remaining lines.  In increasing order of J$_{up}$ from 4 to 13, excluding 6 and 7, the upper limits are 1.5, 1.3, 1.4, 2.6, 3.0, 4.0, 3.0, and 5.2 $\times 10^{-17}$ W m$^{-2}$.

To determine the upper limits, we took an artificial $^{12}$CO input spectrum with zero continuum and all eight lines given fluxes of 
8.7 $\times 10^{-18}$ W m$^{-2}$ and very narrow widths.  We then broadened the lines with a Gaussian 
function having a FWHM of 2300 km s$^{-1}$.  We added the broadened $^{12}$CO line spectrum, multiplied 
by flux scaling factors ranging from 1 to 10, to the observed SLW and SSW subspectra 
in order to judge the flux scaling factor needed for each $^{12}$CO line to yield a definite detection (as 
judged by a line flux measurement that yielded a line flux within 30\% of the input value).

\section{Interpretation}\label{sec:analysis}

Most of the emitting $^{12}$CO is expected to reside in chemically distinct
clumps, composed mostly of oxygen and carbon, in the inner supernova
debris \citep{McCray1993,Jerkstrand2011}. We interpret the observed $^{12}$CO emission lines with a simplified model, in 
which a total mass $M_{CO}$ of $^{12}$CO molecules is found in clumps of uniform 
density and temperature ($T_{CO}$) that occupy a volume $V_{CO}$ of 
the inner debris.  We assume that the inner debris is confined within a 
freely expanding sphere of volume $V_{max}$ defined by radius $R = vt = 2000$ km s$^{-1} \times$ 25 years.  
(Note that the diameter, twice this expansion radius, fully encompasses the FWHM of the line emission).  
The clumps occupy a fraction, $f_{CO}$, of that total volume, such
that $V_{max} \times f_{CO} = V_{CO}$.  In reality, the carbon/oxygen zone is likely composed of
clumps that do vary in chemical composition, density, and temperature \citep{Janka2012}.  Because we do not resolve individual clumps, we 
are modeling the bulk properties that describe the sum of the $^{12}$CO emission.
We calculate the luminosities using 
the Sobolev escape probability approximation \citep{Castor1970}, where the
expressions for the line luminosity, escape probability ($P_{esc}$), and the Sobolev optical
depth ($\tau_S$) take simple forms in the case of freely expanding
supernova debris \citep[equations 5-9]{McCray1993}.  Optically thin emission depends only on 
$M_{CO}$ and $T_{CO}$, whereas at high optical depths, the line luminosity 
is described by the optically thick limit and depends only on $V_{CO}$ and $T_{CO}$.

\begin{figure}
\includegraphics[height=\columnwidth,angle=-90]{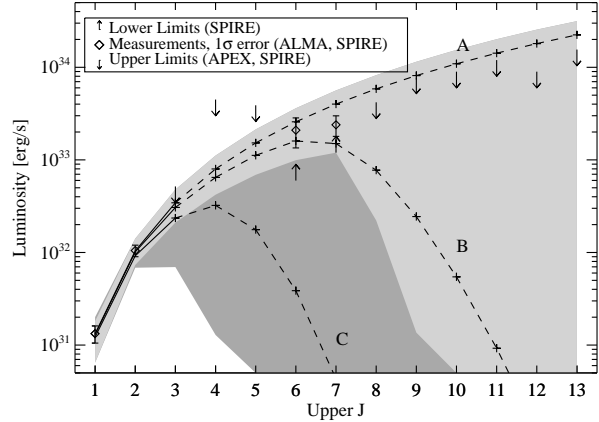}
\caption{$^{12}$CO measurements and upper limits with example spectral line energy distributions (SLEDs).  
Diamonds, downward arrows, and upward arrows (at the base of the arrow) indicate measurements, lower limits, and upper limits, respectively.  
The light gray shaded region illustrates possible LTE SLEDs 
consistent with ALMA measurements and SPIRE lower limits (solutions of at least 5\% of the maximum likelihood solution); the dark gray region indicates additional (low-$T_{CO}$, high-$f_{CO}$) solutions that are excluded by SPIRE lower limits.
Lines A, B, and C are example LTE SLEDs; lines are dashed above J=3-2 to indicate that LTE is an upper limit and the 
true luminosity is likely lower than this curve.  Line A [B, C] corresponds to $T_{CO}$ = 132 [27, 13] K, $M_{CO}$ = 0.55 [0.063,0.028] \ms, $f_{CO}$ = 0.025 [0.141,0.348].  Lines A and B may both be consistent with available data despite arising from very different physical conditions; Line C is an example of a set of conditions that can be excluded given the SPIRE J=6-5 and J=7-5 measurements.
\label{fig:sled}}
\end{figure}

In the above model, we assume that the lines are in Local Thermodynamic Equilibrium (LTE) 
with their surroundings and the populations of the rotational levels are 
described by an excitation temperature equal to the kinetic temperature, $T_{CO}$.
The higher-J transitions are less likely to be in LTE; radiative decays tend to depopulate higher-J rotational levels.  
In general, the population levels and line luminosities will depend on the product of the collision partner density (likely oxygen, carbon, O$_2$ and other $^{12}$CO molecules) and collision rate coefficients, which are unknown for these species.
To make a rough estimate, we examined the collisional rate coefficients for H$_2$ + CO \citep{H2rates}, H + CO \citep{Balakrishnan2002}, and H$_2$O + CO \citep{Green1993}.  The critical densities for the J=2-1 line are one to two orders of magnitude below the modeled density of the O/C clumps \citep[$1.8 \times 10^5$ cm$^{-3}$ at 25 years]{Jerkstrand2011}, indicating that these lines are well approximated by LTE.  The rate cofficients for the proxy species vary significantly for the higher-J lines, introducing considerable uncertainties in modeling the level populations.
Therefore, the LTE predictions for the spectral line energy distributions (SLEDs, as shown in 
Figure \ref{fig:sled}) should be considered upper limits 
for lines above J=3-2; the actual line luminosities for the same physical conditions will be lower.
This means that the current SPIRE upper limits cannot constrain the physical parameters 
of our model, but the J=6-5 and 7-6 detections can exclude parameters that predict low LTE luminosities.

We first consider the fluxes of the $^{12}$CO J=1-0 and 2-1 lines measured by the ALMA, which are in LTE.
We can determine a robust lower limit to the mass by assuming that the J= 1-0 line is optically thin, which yields $M_{CO} > 0.01$ \ms\ at $< 10$ K.
However, the line luminosity ratio $L_{2-1}/L_{1-0} = 7.9 \pm 2.0$ is not consistent with the ratio $>$ 20 expected in the optically thin case.  
Instead, it is consistent with the ratio of 8
expected in the optically thick case ($L_{\rm J,J-1}/L_{1,0} \approx$ J$^3$) in the Rayleigh-Jeans approximation, valid for temperatures above 10 K.  
This case, $L_{thick}$, defines an upper limit to the line
emission because the specific intensity of radiation emitted by a surface
cannot exceed the Planck function, $B_\nu(T_{CO})$.  The $^{12}$CO J=3-2 line is probably also optically thick; assuming so, 
we predict a luminosity of $3.4 \times L_{2-1} = 3.6 \times 10^{32}$ ergs$^{-1}$ and a
peak flux of 65 mJy, lower than the upper limit from APEX.

As mentioned, for a freely expanding volume, the resulting
value of $L_{thick}$ depends only on the net {\it volume}, $V_{CO}$, of the
emitting material and the excitation temperature, independent of $M_{CO}$.  
The product of $T_{CO}$ and $V_{CO}$ is $5.3 \times 10^{52}$ cm$^3$ K.  

Figure \ref{fig:sled} shows the ALMA and SPIRE line luminosities from this work along with example SLEDs (labeled A, B, and C), 
all of which are consistent with the ALMA measurements, despite arising from very different physical conditions.  
We cannot constrain the mass from optically thick emission, but we can
find a minimum mass at which the emission is optically thick and reproduces the ALMA observations.
For all combinations of temperature and $f_{CO} < 1$, the minimum mass is 0.01 \ms\ (3$\sigma$ lower limit).

Though the SPIRE lines cannot be modeled accurately, we can 
rule out any parameter space whose LTE luminosity does not reach the lower limits of the two measured lines.
In Figure \ref{fig:sled}, the lower limits are given by the SPIRE measurements minus 2$\sigma$ (upward arrows). 
The parameter space excluded by the SPIRE lines is highlighted by the dark gray region, clearly excluding 
low-$T_{CO}$ and high-$f_{CO}$ solutions.  
The light gray region shows possible SLEDs allowed by the ALMA, APEX, and SPIRE measurements and 
contains only temperatures greater than 14 K and filling factors less than 0.35; the minimum mass 
to satisfy these conditions is still 0.01 \ms.

\section{Discussion and Conclusions}\label{sec:disc}

From the above analysis, the C/O clumps in 
SN1987A contain at least $0.01$ \ms\ of $^{12}$CO, an order of magnitude greater 
than measured in the first few years after the explosion \citep{Liu1992}. 
The implication is that $^{12}$CO has continued to form over the
past 25 years.
According to a model of the evolution of the
optical and near-IR emission line spectrum \citep{Jerkstrand2011} at $t = 8$
years, the carbon/oxygen
clumps have a mass of carbon plus oxygen of 0.58 \ms\ and a filling
factor $f_{C/O} = 0.02$. Such a small filling factor would imply a higher $^{12}$CO
temperature ($> 100$ K) and at least 0.2 \ms\ of $^{12}$CO (1$\sigma$ lower limit).  

Other studies have specifically examined molecule formation within the distinct chemical zones of the ejecta.  
For example, a 15 \ms\ progenitor at solar metallicity may form $> 0.2$ \ms\ of $^{12}$CO, a significant amount of which 
is formed as early as 600 days after the explosion \citep{Cherchneff2011}.
As the progenitor of SN1987A is most likely a 19 $\pm$ 3 \ms\ blue giant with metallicity similar to the Large Magellanic Cloud 
\citep[1/3-1/4 solar]{Woosley1988}, we can also compare to results from Population III stars (with zero metallicity).  
\citet{Cherchneff2009} found similar results for a Pop III 20 \ms\ progenitor, estimating 0.27 \ms\ of $^{12}$CO produced on the same timescale. 
The velocity adopted for the CO model presented here matches that used in the \citet{Cherchneff2009} study (2242 km s$^{-1}$).   
Though our observations alone can provide only a minimum mass, the combination of the ALMA spectra 
results with the model computed by \citet{Jerkstrand2011} is consistent with theoretical models of molecule formation that have been untested until now.

In addition to the lower limits for temperature and mass, {\it Herschel}-SPIRE provides an upper limit to the total $^{12}$CO luminosity.  
Recently, using the photometric capability of {\it Herschel} (100-500 $\mu$m), the far-IR luminosity of the debris of SN1987A was measured to be 8.45 $\times 10^{35}$ erg s$^{-1}$, and was attributed to thermal emission radiated by a massive reservoir of cold dust grains in the ejecta by \citet{Matsuura2011}.
The lack of detection with SPIRE FTS in the same wavebands provides an upper limit to the contribution from rotationally excited $^{12}$CO molecules to the Far-IR luminosity, with $^{12}$CO contributing at most 8.4\%.  This demonstrates that emission from dust grains is still the most viable explanation for the excess FIR emission detected with {\it Herschel}.

We have presented measurements for four $^{12}$CO rotational transitions in the ejecta of SN1987A, 
as well as upper limits for eight additional lines.  The unprecedented angular resolution of ALMA 
firmly locates this emission as originating from the inner ejecta.  The J=2-1 and J=1-0 lines are 
optically thick and require at least 0.01 \ms\ of $^{12}$CO.  These results, combined with previous dust results from {\it Herschel}, illustrate 
a SN environment filled with cool molecules and gas merely 25 years after the explosion.  
These results are not only the first of their kind, they are also exciting as a demonstration for 
future work.  Unlike optical observations, where the redshifted emission is largely obscured by dust, (sub)millimeter observations can see the entire velocity range of the debris.  Surfaces of constant Doppler shift are planar sections of the freely expanding SN debris, and the $^{12}$CO lines are sufficiently bright that we will be able to image the debris in 3 dimensions with resolution better than 0.02\arcsec\ through Doppler tomography with the full ALMA array \citep{Isensee2010,DeLaney2010,Larsson2013}.  
ALMA will be able to do the same with different molecules (e.g. SiO) to probe chemically distinct regions, which likely have different morphologies.

\acknowledgements
The National Radio Astronomy Observatory is a facility of the National Science Foundation operated under cooperative agreement by Associated Universities, Inc.
This paper makes use of the following ALMA data: ADS/JAO.ALMA\#2011.0.00273.S. ALMA is a partnership of ESO (representing its member states), NSF (USA) and NINS (Japan), together with NRC (Canada) and NSC and ASIAA (Taiwan), in cooperation with the Republic of Chile. The Joint ALMA Observatory is operated by ESO, AUI/NRAO and NAOJ.
J.K. acknowledges funding from the National Science Foundation Graduate Research Fellowship Program.  
R.K.'s supernova research at Harvard is supported by the NSF through grant AST-1211196.  
J.B., P.v.H., and P.R.  acknowledge support from the Belgian Science Policy office through the ESA Prodex program. 
M.L. acknowledges an ESO/Keele studentship.
J.M. received partial support from grants AYA2009-13036-C02-02 of the MICINN and PROMETEO 104/2009 of the Generalitat Valenciana.
M.Meixner acknowledges funding support from NASA {\it Herschel} Space Center, JPL contracts \#1381522 and 1381650 and NASA NAG5-12595.

{\it Facilities:} \facility{ALMA}, \facility{Herschel}, \facility{APEX}.


\end{document}